\documentclass[a4paper,11pt]{article}
\pdfoutput=1

\usepackage{jheppub}

\usepackage{subfig}
\usepackage{xspace}
\usepackage[countmax]{subfloat}
\usepackage{slashed}

\usepackage{comment}

\usepackage{bbm}

\newcommand{\nbar}{{\bar n}}

\def\be{\begin{equation}}
\def\ee{\end{equation}}

\def\zcut{z_{\text{cut}}}

\def\nbar{\bar n}

\DeclareRobustCommand{\Eq}[1]{Eq.~(\ref{#1})}

\DeclareRobustCommand{\Ref}[1]{Ref.~\cite{#1}}
\DeclareRobustCommand{\Refs}[1]{Refs.~\cite{#1}}

\usepackage{color}
\definecolor{darkblue}{rgb}{0,0,0.5}
\definecolor{darkred}{rgb}{0.5,0,0}
\definecolor{purple}{rgb}{0.5,0.0,0.5}

\title{Non-Global and Clustering Effects for Groomed Multi-Prong Jet Shapes}

\author[1]{Duff Neill}
\affiliation[1]{Theoretical Division, MS B283, Los Alamos National Laboratory, Los Alamos, NM 87545, USA}

\emailAdd{duff.neill@gmail.com}

\abstract{We present a resummation of the non-global and clustering effects in groomed (with modified mass drop tagger) multi-pronged observables, valid to next-to leading logarithmic accuracy in the $D_2$ distribution (all single logarithmic terms), focusing on the non-global and clustering effects which cannot be removed by normalizing the cross-section. These effects are universal in the sense that they depend only on the flavor structure of the $1\to 2$ splitting forming the multi-pronged subjets and the opening angle of the splitting, being insensitive to the underlying hard process or underlying event. The differential spectra with and without the non-global and clustering effects are presented, and the change in the spectra is found to be small.}

\begin{document}
\maketitle

\section{Brief Discussion of Clustering Effects and NGLs}
In Refs. \cite{Larkoski:2017cqq} and \cite{Larkoski:2017iuy}, a factorization for two versus one pronged jets using the jet shape $D_2$ and the modified mass drop tagger (mMDT) \cite{Dasgupta:2013ihk,Dasgupta:2013via} or equivalently, soft drop with angular exponent $\beta = 0$  \cite{Larkoski:2014wba} was presented. Within \Ref{Larkoski:2017cqq}, it was argued that the non-global and clustering effects do exist (in a limited sense) when the jet has genuinely two prongs. The two prongs creates an effective jet area for secondary splittings to radiate into, where the groomer cannot remove them. The purpose of this note is to provide numerical evidence that these effects are small, by detailing the large-$N_c$ resummation to leading logarithmic accuracy of both the clustering and non-global effects, justifying the claimed next-to-leading logarithmic (NLL) accuracy of the resummation in \Ref{Larkoski:2017cqq}, that is, all terms of the form $\alpha_s\text{ln}D_2$. An outline of the algorithm used here to estimate the effects of the non-global radiation was already presented in \Ref{Larkoski:2017cqq}, where the numerical results were used to conclude that the NGL effects were small. Here we simply provide a detailed analysis of the numerics and presentation of the algorithm.\footnote{For a detailed discussion of the physical observable, its factorization and resummation, we refer the reader to \Refs{Larkoski:2017cqq} and also \cite{Larkoski:2015kga}.}\footnote{A similar algorithm for calculating the contribution of non-global logarithms was investigated in Refs. \cite{Appleby:2002ke,Banfi:2005gj}. There the algorithm focused on soft radiation in gaps-between-jets, where all radiation was organized via the $K_T$-clustering algorithm, rather than the mMDT grooming algorithm. These works encountered similar clustering effects modifying the structure of the jet boundary, though the precise details of how the clustering history was treated in the algorithm were lacking.} 

The clustering and non-global effects\footnote{For recent work on the theory of non-global correlations, see \cite{Hatta:2008st,Avsar:2009yb,Hatta:2013iba,Caron-Huot:2015bja,Larkoski:2015zka,Neill:2015nya,Becher:2015hka,Becher:2016mmh,Larkoski:2016zzc,Neill:2016stq,Hatta:2017fwr,Becher:2016omr,Becher:2017nof,Balsiger:2018ezi}.} are incorporated formally in our factorization formulae within the collinear-soft function:
\begin{align}\label{eq:collinear_soft_function}
C_s\Big(e_{3},z_{\text{cut}}\Big)&=\frac{1}{N_i}\text{tr}[\mathbf{T}\langle 0|T\{S_{a}S_{b}S_{\bar{n}}\}\Theta\Big(e_{3}-\mathbf{E}_3\big|_{SD}\Big)\Theta_{SD}(z_{\text{cut}})\bar T\{S_{a}S_{b}S_{\bar{n}}\}|0\rangle\mathbf{T}]\,.
\end{align}
$\bar{n}=(1,-\hat{n})$ is the recoiling direction of the jet, whereas the directions $a=(1,\hat{a})$ and $b=(1,\hat{b})$ are the directions of legs of the dipole structure selected by the requirement $e_{3}\ll (e_2)^3$, that is, the directions of the subjets, with some opening angle $\theta_{ab}$. $\mathbf{T}$ is a color generator fixed by the flavor structure of the splitting, that the color indicies of the wilson lines contract into. The jet axis is $\hat{n}=\frac{\hat{a}+\hat{b}}{|\hat{a}+\hat{b}|}$. Thus the soft function is sensitive to two soft scales set by $e_{3}$, and $z_{\text{cut}}$, as well as the modified geometrical structure of the jet due to multiple emissions. Emissions that are clustered into legs $a$ and $b$ at one emission level may not be so clustered at higher orders due to other emissions that are closer in angle, yet outside the clustering region of the two-prongs. The boundary set by $R$ of the groomed jet is irrelevant for the issue of non-global logarithms (NGL) of the groomed jet distribution \emph{for a specific flavor of jet}, since we may always take the emissions that do not cluster into the dipole to fail soft drop, wherever they are. However, in hadron-hadron collisions, there are non-global contributions to the quark and gluon fractions which can be sensitive to the boundary of the jet. Thus the only relevant geometrical constraint is that we are concerned about the history of emissions at angles greater than $\theta_{ab}$ to $\hat{n}$, where the precise region that is clustered into the legs is given by the Cambridge/Aachen algorithm \cite{Dokshitzer:1997in,Wobisch:1998wt,Wobisch:2000dk}.

We emphasize at this level already, the NGLs and clustering logs within the groomed jet are much less worrisome than traditional NGLs. This is due to the fact that the NGLs and clustering logs are local to the jet, that is, process independent, and largely determined by the flavor structure of the splitting giving rise to the subjets, and possibly the opening angle of the subjets $\theta_{ab}$. This is due to the fact that the subjet splitting is boosted deep into the fat jet, so all other eikonal lines of the process are collapsed onto a single recoiling direction.\footnote{This feature of NGL distributions in a boosted jet was already noted in \Ref{Dasgupta:2012hg}, where it was argued that the NGL's for the ungroomed jet mass distribution at small cone size $R\ll 1$ are the same as the NGL's for the hemisphere jet mass distributions, see also \Ref{Schwartz:2014wha} for an argument based on conformal symmetry.} The grooming algorithm guarantees only these quasi-collinear eikonal lines can contribute to the NGL distribution, and the possible clustering effects. That is, if a set of eikonal lines emits into the region subtended by the subjet splitting, and neither leg of the radiating dipole is the $a$ or $b$ leg of the subjets, the probability for the emission to be clustered will be proportional to $\theta_{ab}^2\ll 1$, a power suppressed contribution. Thus only the boosted set of eikonal lines forming the subjet splitting can contribute. Hence the distribution is universal, and can in principle be computed once and for all. 

We give a procedure to compute these effects below, which we used to estimate whether our predictions for the distributions from exponentiating the global contributions were sufficient for NLL acurracy. The main theoretical interest is the necessity to keep track of the whole history of emissions in the cascade of soft partons, due to the Cambridge/Aachen clustering metric.\footnote{This would complicate an evolution equation description of the kind found in \Ref{Banfi:2002hw} and \cite{Weigert:2003mm}.} This entangles the emissions off of distinct dipoles, which can often be considered independently in their evolution history if the phase-space constraint is geometric to all orders in perturbation theory.

\section{Review of Groomed Jet Shapes}
\begin{figure}\centering
\includegraphics[scale=0.6]{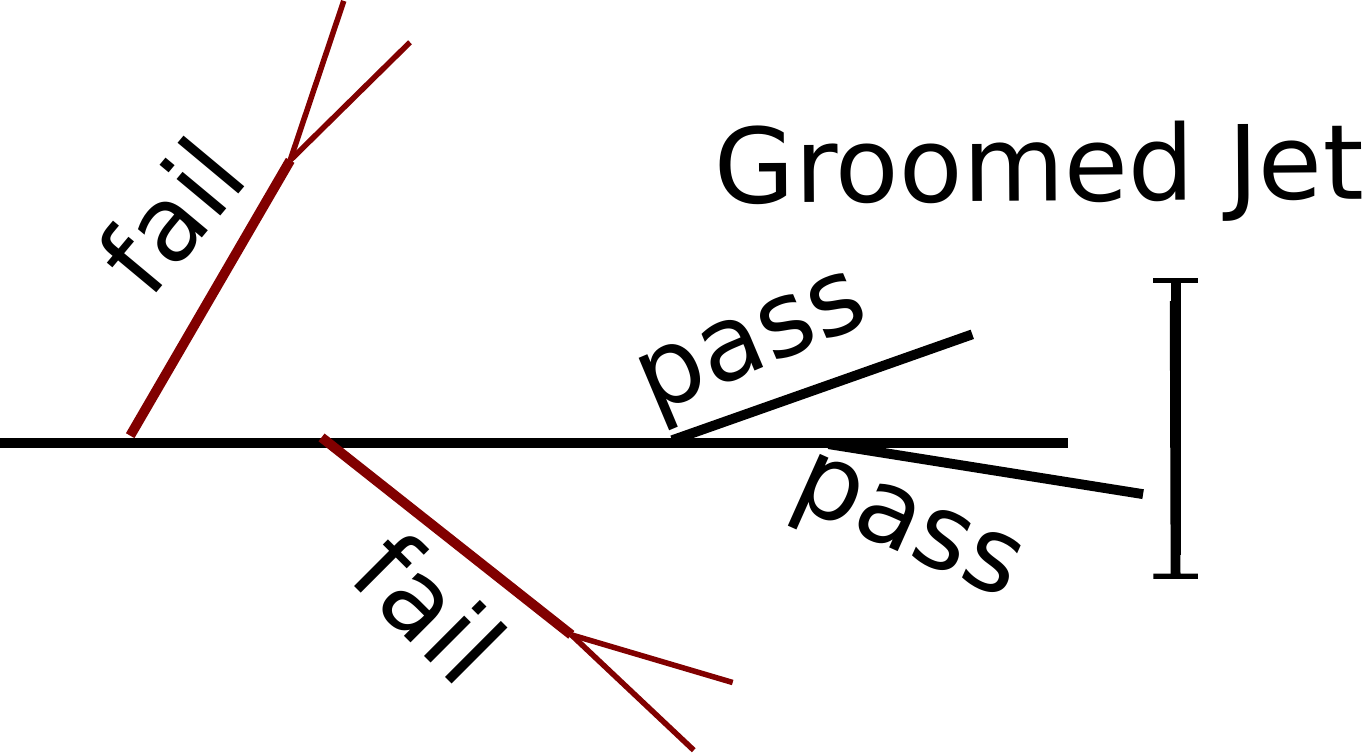}
\caption{\label{fig:mMDT} Pictorial representation of the action of the mMDT on a candidate jet.}
\end{figure}
For completeness, we give a review of the mMDT procedure that underpins much of the analytical progress in understanding jet substructure, as well as the the $D_2$ jet shape observable for distinguishing 1 versus 2 pronged jet substructure. The mMDT procedure grooms a candidate jet, which could have been constructed out of the event from any other suitable jet algorithm, by reclustering the constituents with the C/A algorithm. At each step in the algorithm, the two particles (the daughters) whose momenta have the smallest opening angle are combined into a pseudo-particle (the parent) with the same total energy, and pointing (typically) either in the direction of the sum of the momenta (called the E-scheme), or in the direction of the more energetic of the two particles being combined (the winner-takes-all scheme). The two particles being combined are deleted from the list of particles in the jet and are replaced by the pseudo-particle. This reclustering continues until all particles are combined into a single pseudo-particle. The clustering tree then is the history of the recombinations. 

To groom the jet, the clustering tree is examined in the opposite order that it was constructed. One takes the current pseudo-particle (starting with the psuedo-particle forming the total jet, the last clustering of the C/A algorithm) and examines the two daughters that compose it. Let $z_i$ and $z_j$ be the energy fractions of the two daughters. The current pseudo particle is declared to be the groomed jet if:
\begin{align}\label{eq:groomed_success}
\frac{\text{min}\{z_i,z_j\}}{z_i+z_j}>\zcut\,.
\end{align}
Where the parameter $\zcut$ is an input to the grooming procedure. If the two daughters fail the condition of Eq. \eqref{eq:groomed_success}, then the less energetic of the two are discarded from the jet, and the more energetic daughter is now the current psuedo-particle to be declustered and tested. A pictorial representation of the mMDT can be found in Fig. \ref{fig:mMDT}. The solid black line follows the most energetic branch, and the two red branches are at the widest angle, and so are the first to be declustered. They are assumed to fail the mMDT criterion, and further down the clustering tree of the most energetic line, we find the branching which passes mMDT at a much smaller angle.

Once we have found the branching which passes the mMDT condition of Eq. \eqref{eq:groomed_success}, all the particles which form the two branches are the groomed jet, and any measurement on those particles are a groomed measurement. For the purposes of multi-prong discrimination, that is, finding whether a jet has a two subjets or a single hard core, a convinient shape variable is the so-called $D_2$ observable of Ref. \cite{Larkoski:2014gra}, which is formed from the ratio of two energy-energy correlation measurements of Ref. \cite{Larkoski:2013eya}. Let $J$ be the list of particles within the jet, and $G(J,\zcut)$ the list of particles that survive the mMDT grooming procedure. Then the appropriate shape variable is given by:
\begin{align}
D_2^{(\beta)}&=\frac{e_3^{(\beta)}}{\Big(e_2^{(\beta)}\Big)^3}\,,\\
e_2^{(\beta)}&=\frac{1}{2}\sum_{i,j\in G(J,\zcut)}\frac{E_i}{E_J}\frac{E_j}{E_J}\Big(\frac{2 p_i\cdot p_j}{E_i E_j}\Big)^{\frac{\beta}{2}}\,,\\
e_3^{(\beta)}&=\frac{1}{6}\sum_{i,j,k\in G(J,\zcut)}\frac{E_i}{E_J}\frac{E_j}{E_J}\frac{E_k}{E_J}\Big(\frac{2 p_i\cdot p_j}{E_i E_j}\frac{2 p_j\cdot p_k}{E_j E_k}\frac{2 p_k\cdot p_i}{E_k E_i}\Big)^{\frac{\beta}{2}}\,.
\end{align}
$E_J$ is the total jet energy, $p_i$ is the four momenta of the $i$-th particle, and $E_i$ is its energy. When $\beta=2$, we will simply write $D_2^{(2)}=D_2$. When $D_2\ll 1$ we have a jet with genuinely two distinct subjets.

For $e^+e^-\rightarrow\, hadrons$, the cross-section for a groomed jet in the limit $D_2\ll 1$ when the two subjets are collinear assumes the factorized form:
{\small\begin{align}\label{eq:factorized_cross_section}
\frac{d\sigma}{dz de_{2}^{(\beta)} de_{3}^{(\beta)}}&=\sigma_0 H\Big(Q^2\Big)J_{\bar{n}}\Big(Q^2\Big)S\Big(Q\zcut,R\Big)H_{1\rightarrow 2}\Big(z,e_{2}^{(\beta)}\Big)J_{1}(e_{3}^{(\beta)})\otimes J_{2}(e_{3}^{(\beta)})\otimes C_{s}\Big(e_{3}^{\beta},\zcut\Big)+...\,.
\end{align}}
Within this factorization, $Q$ is the center of mass energy, $H$ is the hard matching coefficient describing the decay of a off-shell photon or $Z$-boson into a quark/anti-quark pair at the scale $Q$, or a Higgs boson into gluons, $J_{\bar{n}}$ is the recoiling jet function, $S$ is the global soft function describing the ungroomed jet boundaries and the structure of the soft radiation which fails mMDT, $H_{1\rightarrow 2}$ describes the formation of the two collinear subjets, labeled $1$ and $2$, with energy fractions $z$ and $1-z$ (relative to $Q/2$, the energy of the fat parent jet), $J_1$ and $J_2$ are jet functions describing the collinear contributions to the $e_{3}^{(\beta)}$ measurement, and $C_{s}$ is the collinear soft-function of Eq. \eqref{eq:collinear_soft_function}, describing the soft radiation which is sensitive to the dipole structure of the subjets. The function $H_{1\rightarrow 2}$ is given by the finite terms of the $1\rightarrow 2$ squared collinear splitting amplitudes. Field theoretic definitions of all functions, their calculation to one-loop accuracy, their renormalization, and (global) resummation can be found in Ref. \cite{Larkoski:2017cqq}, as well as factorizations for the regions where a subjet becomes soft-collinear, or when $D_2\sim 1$. In this paper we focus on the non-global resummation of $C_{s}$.

At $\beta=2$, we can compute the $D_2$ distribution with a cut on the mass by the marginalizing over the cross-section:
\begin{align}
  \frac{d\sigma}{dD_2}(m_{\text{min}},m_{\text{max}})&=\int_0^1 dz\int_{0}^{\infty} de_{3}\int\displaylimits_{m^2_{min}/Q^2}^{m^2_{max}/Q^2}de_2\,\delta\Big(D_2-\frac{e_3}{(e_2)^3}\Big)\frac{d\sigma}{dz de_{2} de_{3}}\,.
\end{align}  

\section{Monte Carlo Algorithm for Clustering Logs and NGLs}
This is a description of a Monte Carlo algorithm for computing the soft drop non-global logarithms and clustering effects for the $e_3$ distribution. The main idea was summarized in \Ref{Larkoski:2017cqq}, Appendix E, and follows the scheme first outlined in \Ref{Dasgupta:2001sh}, but we now spell out the algorithm in detail. To single-logarithmic accuracy in the large $N_c$ limit, this algorithm computes the NGLs and clustering logs for the collinear soft function $C_s$ given above. We need:
{\small\begin{itemize}
\item All four vectors are null, and determined by their spatial direction, being of the form: $a=(1,\hat{a})$. Thus for all dot products between two null vectors, we have $a\cdot b= 1- \cos \theta_{ab}$, with $\theta_{ab}$ being the angle between them in the laboratory frame.
\item $a,b$ are the hard prongs of the soft dropped jet.
\item List of emissions $E$.
\item List of dipoles $\mathcal{D}$, where an element is given by $\{x,y,\Delta\eta_{xy}\}$. $x,y$ are the null directions of the legs, and $\Delta\eta_{xy}$ is their opening angle in rapidity. 
\item Histogram $H_t$, indexed by resummation time $t=\frac{\alpha_s C_A}{\pi}\text{ln}\frac{z_{cut}}{e_3}$.
\item Angular cutoff $\delta$.
\item $w$, the weight of the current event.
\item Rapidities are calculated in the lab frame.
\end{itemize}}
In what follows, we denote the angle between eikonal lines $x$ and $y$ by $\theta_{xy}$. We let:
\begin{align}
W_{x y}(j)&=\frac{x\cdot y}{(x\cdot j)\,(j\cdot y)}\,,\\
\Delta\eta_{xy}&=\int\frac{d\Omega_j}{4\pi}\theta(1-\text{cos}\delta-x\cdot j)\theta(1-\text{cos}\delta-y\cdot j)W_{x y}(j)\,.
\end{align}
First zero out the histogram $H_t=0 \,,\forall t$.
\begin{itemize}
\item {\bf Initialize:}\begin{align}
E=\{a,b\}\,, & & \mathcal{D}=\Big\{\{a_1,b_1,\Delta\eta_{a_1 b_1}\},\{a_2,b_2,\Delta\eta_{a_2 b_2}\},...,\{a_n,b_n,\Delta\eta_{a_n b_n}\}\Big\}\,,  && t=0\,, && w = 1\,.
\end{align}
\end{itemize}
For instance, for the various splittings, we have the dipole structures:
\begin{align}
g\rightarrow gg:&\qquad \mathcal{D}=\Big\{\{a,b,\Delta\eta_{a b}\},\{a,\nbar,\Delta\eta_{a\nbar}\},\{\nbar,b,\Delta\eta_{b\nbar}\}\Big\}\,,\\
q\rightarrow qg:&\qquad \mathcal{D}=\Big\{\{a,b,\Delta\eta_{a b}\},\{a,\nbar,\Delta\eta_{a\nbar}\}\Big\}\,,\\
g\rightarrow q\bar{q}:&\qquad \mathcal{D}=\Big\{\{a,\nbar,\Delta\eta_{a\nbar}\},\{\nbar,b,\Delta\eta_{b\nbar}\}\Big\}\,,\\
Z\rightarrow q\bar{q}:&\qquad \mathcal{D}=\Big\{\{a,b,\Delta\eta_{ab}\}\,.
\end{align} 
The differential probability for generating an emission in direction $j$ from the set of dipoles ${\mathcal D}$ is then:
\begin{align}
\tilde{F}_{\mathcal{D}}(j)&=\sum_{i\in \mathcal{D}}W_{x_iy_i}(j)-C_{ab}\Big(j,\{a,b\}\Big)W_{ab}(j)
\end{align}
Where the soft drop virtual subtraction phase-space $C_{ab}\Big(j,\{a,b\}\Big)$ is defined in \Eq{eq:SD_phase_space}. The algorithm is now:
{\small\begin{enumerate}
\item Calculate $\Delta\eta_{tot}$ by summing over the $\Delta\eta$'s in $\mathcal{D}$. Generate a random $\Delta t$ via the probability distribution:
  \begin{align}
    P(\Delta t)&\propto \text{exp}\Big(-\Delta\eta_{tot}\Delta t\Big)
  \end{align}
Increase $t$ by $\Delta t$.
\item Select the dipole $\{x,y,\Delta\eta_{xy}\}\in \mathcal{D}$ randomly with probability $\frac{\Delta\eta_{xy}}{\Delta\eta_{tot}}$.
\item Create emission $j$ randomly with distribution $W_{xy}(j)$, such that $x\cdot j,y\cdot j > 2\text{sin}^2\frac{\delta}{2}$.
\item Calculate:
  \begin{align}\label{eq:SD_phase_space}
    C_{ab}\Big(j,\{a,b\}\Big)&=\theta\Big(\theta_{ab}-\theta_{aj}\Big)\theta\Big(\theta_{bj}-\theta_{aj}\Big)+\theta\Big(\theta_{ab}-\theta_{bj}\Big)\theta\Big(\theta_{aj}-\theta_{bj}\Big)\,\nonumber\\
    C_{ab}(j,E)&=\theta\Big(\theta_{ab}-\theta_{aj}\Big)\prod_{i\in E}\theta\Big(\theta_{ij}-\theta_{aj}\Big)+\theta\Big(\theta_{ab}-\theta_{bj}\Big)\prod_{i\in E}\theta\Big(\theta_{ij}-\theta_{bj}\Big)\,
  \end{align}
where $\theta_{xy}$ is the angle between $x,y$.
\item If $C_{ab}(j,E)=0\, \&\, C_{ab}\Big(j,\{a,b\}\Big)=0$, then the emission is not clustered into the emissions $a$ or $b$ before $a,b$ are clustered together, nor is it possibly a virtual subtraction. Delete $\{x,y,\Delta\eta_{xy}\}$ from $\mathcal{D}$, add $\{x,j,\Delta\eta_{xj}\}$ and $\{j,y,\Delta\eta_{jy}\}$ to $\mathcal{D}$, and add $j$ to $E$. Goto step 1. 
\item If $C_{ab}(j,E)=0\, \&\, C_{ab}\Big(j,\{a,b\}\Big)>0$, then the emission is not clustered into the emissions $a$ or $b$ before $a,b$ are clustered together, but it can be a virtual subtraction. 
    \begin{itemize}
    \item $x \neq a_i$ and $y \neq b_i$ for all original dipoles $\{a_i,b_i\}$. Delete $\{x,y,\Delta\eta_{xy}\}$ from $\mathcal{D}$, add $\{x,j,\Delta\eta_{xj}\}$ and $\{j,y,\Delta\eta_{jy}\}$ to $\mathcal{D}$, and add $j$ to $E$. Goto step 1. 
    \item Either $x$ or $y$ is a leg of an original dipole. Then $X =\, ${\bf ComputeVeto}$(j,x,y)$.
    \item If $X>0$, then with probability $X$, we delete $\{x,y,\Delta\eta_{xy}\}$ from $\mathcal{D}$, add $\{x,j,\Delta\eta_{xj}\}$ and $\{j,y,\Delta\eta_{jy}\}$ to $\mathcal{D}$, and add $j$ to $E$. Goto step 1.  Otherwise, we throw away this emission $j$, and goto step 1.
    \item If $X\leq 0$, then add $wX$ to $H_t$, reset $w$ to $w(1-X)$, and goto step 1.
    \end{itemize}
\item If $C_{ab}(j,E)>0$, then we cluster $j$ into $a$ or $b$, then 
    \begin{itemize}
    \item $x \neq a_i$ and $y \neq b_i$ for all original dipoles $\{a_i,b_i\}$. Then add $w$ to $H_t$ and start new event, re-initialize.
    \item If either $x$ or $y$ are legs of the original dipole, $X = \,${\bf ComputeVeto}$(j,x,y)$.
    \item If $X>0$, then with probability $X$, we add $w$ to $H_t$ and start a new event, re-initialize. Otherwise, we throw away this emission $j$, and goto 1.
    \item If $X\leq 0$, then add $wX$ to $H_t$, reset $w$ to $w(1-X)$, and goto step 1.
    \end{itemize}
\end{enumerate}}

The emissions are generated in step (3) in the selected dipole rest frame, where they can be given by generating uniform distributions of vectors in rapidity and azimuth with respect to the back-to-back eikonal lines. We then boost back to the lab-frame and check the angular cutoff conditions are satisfied.

\subsubsection*{The Veto}
This is how we compute the reweighting veto, {\bf ComputeVeto}$(j,x,y)$ :
{\small\begin{itemize}
    \item $x=a_i, y=b_i$, that is, $\{x,y\}$ is one of the original dipoles, then:
      \begin{align}
        X=0
      \end{align}
    \item $x=a_i$, a leg of one of the original dipoles $\mathcal{D}_i$, but $y\neq b_i$, then, 
        \begin{align}
          X&=1-\frac{W_{a_i b_i}(j)}{W_{a_i y}(j)}\theta\Big(\theta_{b_i j}-\theta_{a_i j}\Big)\label{eq:subtraction_a_leg}
        \end{align}
    \item $y=b_i$, a leg of one of the original dipoles $\mathcal{D}_i$, but $x\neq a_i$, then, 
        \begin{align}
          X&=1-\frac{W_{a_i b_i}(j)}{W_{x b_i}(j)}\theta\Big(\theta_{a_i j}-\theta_{b_i j}\Big)\label{eq:subtraction_b_leg}
        \end{align}
    \item Return $X$.

\end{itemize}}
The calculation of the reweighting $X$ value in the {\bf ComputeVeto} function splits the virtual subtraction between the two legs of an initial dipole according to which leg it is closer to. So for instance if the virtual subtraction is due to the initial dipole which is the $a,b$-dipole forming the subjets, we justify the partitioning of the subtraction as follows:
\begin{align}
\tilde{F}_{\mathcal{D}}(J)&=\sum_{i\in \mathcal{D}}W_{x_iy_i}(j)-W_{ab}(j)\theta_{SD}\\
&=\sum_{i\in \mathcal{D}/{a\text{ or }b}_{dip}}W_{x_iy_i}(j)+W_{ay}(j)+W_{zb}(j) -W_{ab}(j)\theta_{SD}\\
&=\sum_{i\in \mathcal{D}/{a\text{ or }b}_{dip}}W_{x_iy_i}(j)+\Bigg(W_{ay}(j) -W_{ab}(j)\theta\Big(\theta_{b j}-\theta_{a j}\Big)\theta_{SD}\Bigg)+\Bigg(W_{zb}(j) -W_{ab}(j)\theta\Big(\theta_{a j}-\theta_{b j}\Big)\theta_{SD}\Bigg)
\end{align}
where the virtual subtraction has angular phase space given by $\theta_{SD}=C_{ab}\Big(j,\{a,b\}\Big)$, see \Ref{Larkoski:2017cqq}. We note that the phase-space given by the function $C_{ab}\Big(j,\{a,b\}\Big)$ is the same angular phase-space used to define the (sudakov) global logarithms. If the virtual subtraction is from an $a-\bar{n}$ or $b-\bar{n}$-dipole, that is, a dipole formed from an initial leg and the recoiling direction, then the $\theta$-function in Eq. \eqref{eq:subtraction_a_leg} or Eq. \eqref{eq:subtraction_b_leg} is always satisfied. That is, an emission that is closer to the recoil direction then to either leg $a$ or leg $b$ cannot satisfy $\theta_{SD}>0$ when $\theta_{ab}\ll 1$.\footnote{In the shower, we track the origin of the initial legs as they migrate to new dipoles. When performing the veto step, we compute the veto using the original dipole that the leg currently radiating was initially attached to. That is, a radiating dipole attached to the leg $a$ could have had leg $a$ descended from either the $a-b$ dipole and the $a-\bar{n}$-dipole, when leg-$a$ is a gluon. We compute the weight $X$ in the {\bf ComputeVeto} function according to whichever original dipole that the leg is descended from.}

The real emissions have an angular phase space that is dictated by the complete emission history up to this point. Thus the algorithm naturally incorporates the clustering effects that arises from mis-matching phase space constraints between the exponentiated one loop result and the result given by multiple emissions.

\section{Discussion of Cambridge/Aachen Clustering History}
\begin{figure}\centering
\includegraphics[scale=0.6]{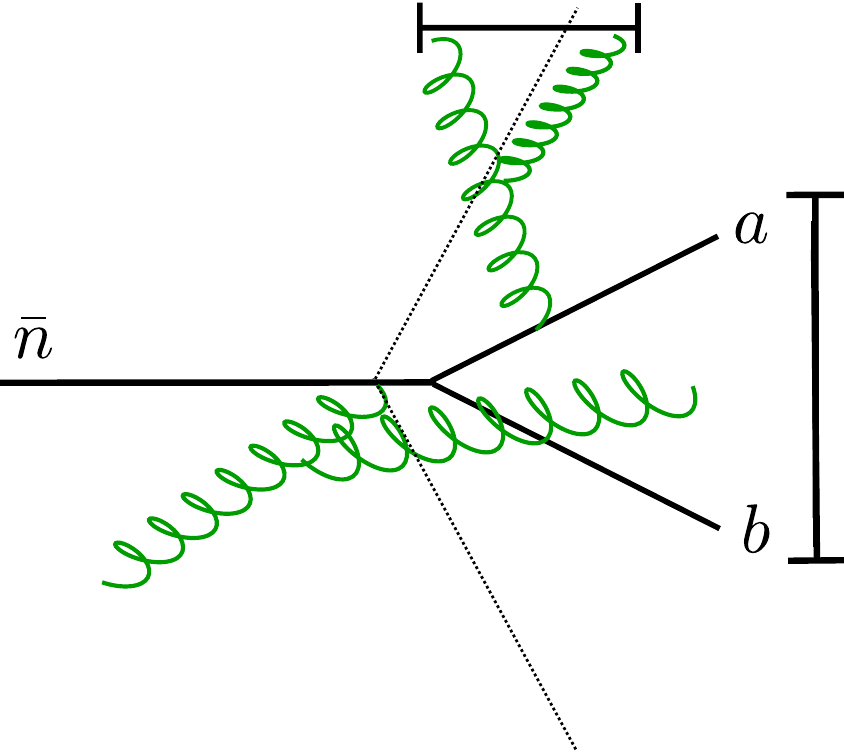}
\caption{\label{fig:ngl_cluster} Illustration of how emissions outside the jet can cluster with emissions within the initial angular region set by the subjet splitting. The initial angular region that sets the resummation of global sudakov effects is given by the dashed lines. Some clustering can remove emissions that would naively contribute to $D_2$.}
\end{figure}

In the above algorithm, we are working in the strongly energy-ordered limit. Formally, every emission has an energy much greater than all subsequent emissions. This is justified in part due to the fact that the collinear-soft function itself is a product of eikonal lines, and thus already contains the strongly energy-ordered QCD diagrams as a proper subset of its full diagrammatic expansion.

Since the emissions are strongly energy-ordered, if emission $p_j$ is produced late in the cascade, we simply need to compare the angle that this emission has to all previous emissions, assuming the previous emissions satisfy:
\begin{itemize}
\item They fail soft drop on their own.
\item They have not clustered into $a$ or $b$ before $a,b$ are themselves clustered.
\end{itemize}
Emission $p_j$ will be clustered into whatever prior emission it is closest to in angle. Moreover, by the strong energy ordering assumption, it will not change the direction of that emission. Thus it can only contribute to the observable $e_3$ if it manages to be clustered into $a$ or $b$ before it is clustered into any emission generated so far in the cascade, that is whether:
  \begin{align}
C_{ab}(j,E)>0 &\text{ or } C_{ab}(j,E)=0\,,\\
    C_{ab}(j,E)&=\theta\Big(\theta_{ab}-\theta_{aj}\Big)\prod_{i\in E}\theta\Big(\theta_{ij}-\theta_{aj}\Big)+\theta\Big(\theta_{ab}-\theta_{bj}\Big)\prod_{i\in E}\theta\Big(\theta_{ij}-\theta_{bj}\Big)\,.
  \end{align}
Note that any later emission after a given real emission has been established in the cascade cannot change the directions of the emissions it may be clustered into: the resulting pseudo-particle in C/A will point in the direction of the more energetic emission. This is exactly true if one used a winner-take-all clustering scheme of Refs. \cite{Bertolini:2013iqa,Larkoski:2014uqa}, and approximately true if using a standard E-scheme, where one simply sums the momentum. Thus softer emissions cannot change whether $p_j$ above is clustered into $\{a,b\}$ or not. The action of the clustering history is illustrated in Fig. \ref{fig:ngl_cluster}.

\section{Numerical Results}
\begin{figure}\centering
\includegraphics[scale=0.45]{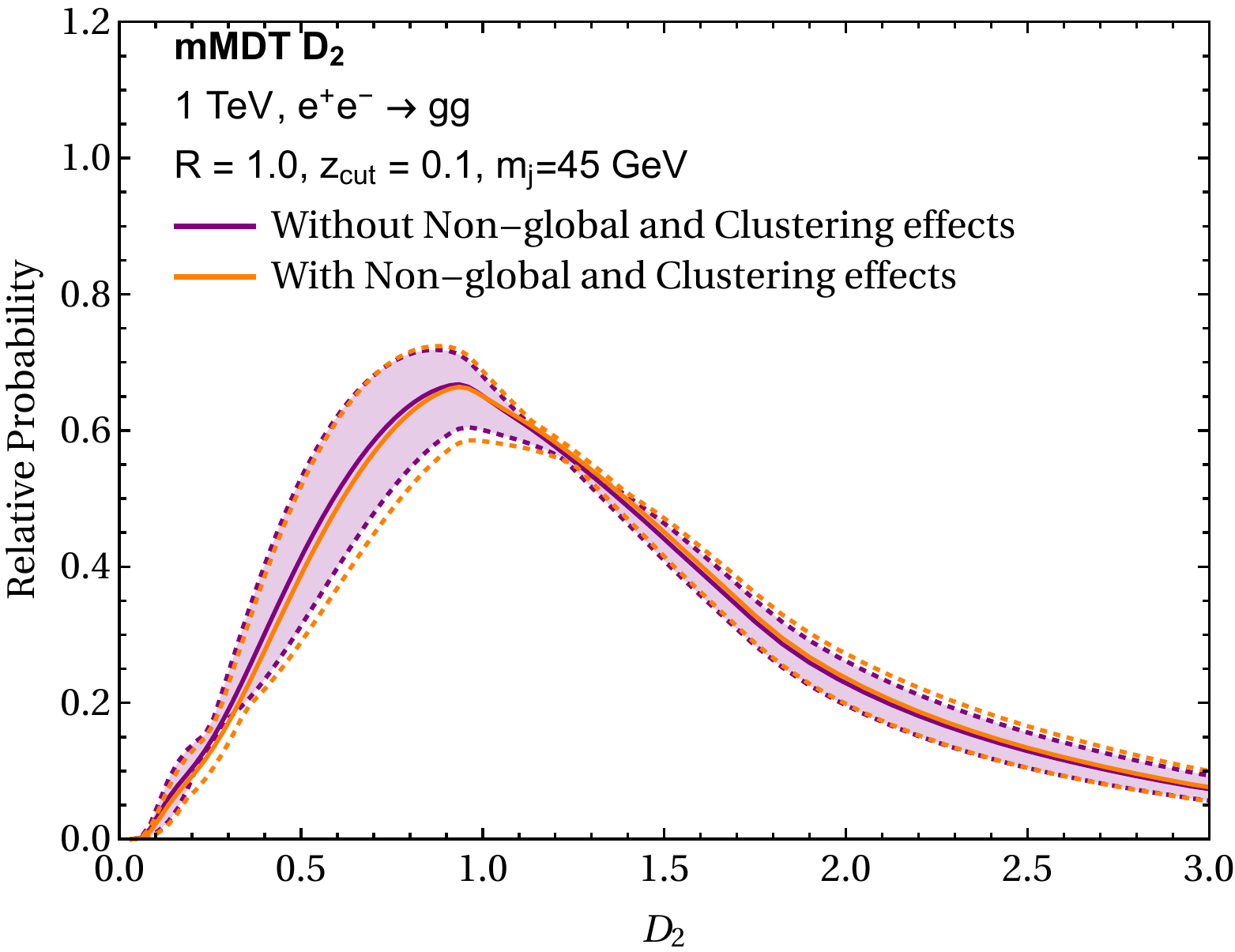} 
\includegraphics[scale=0.45]{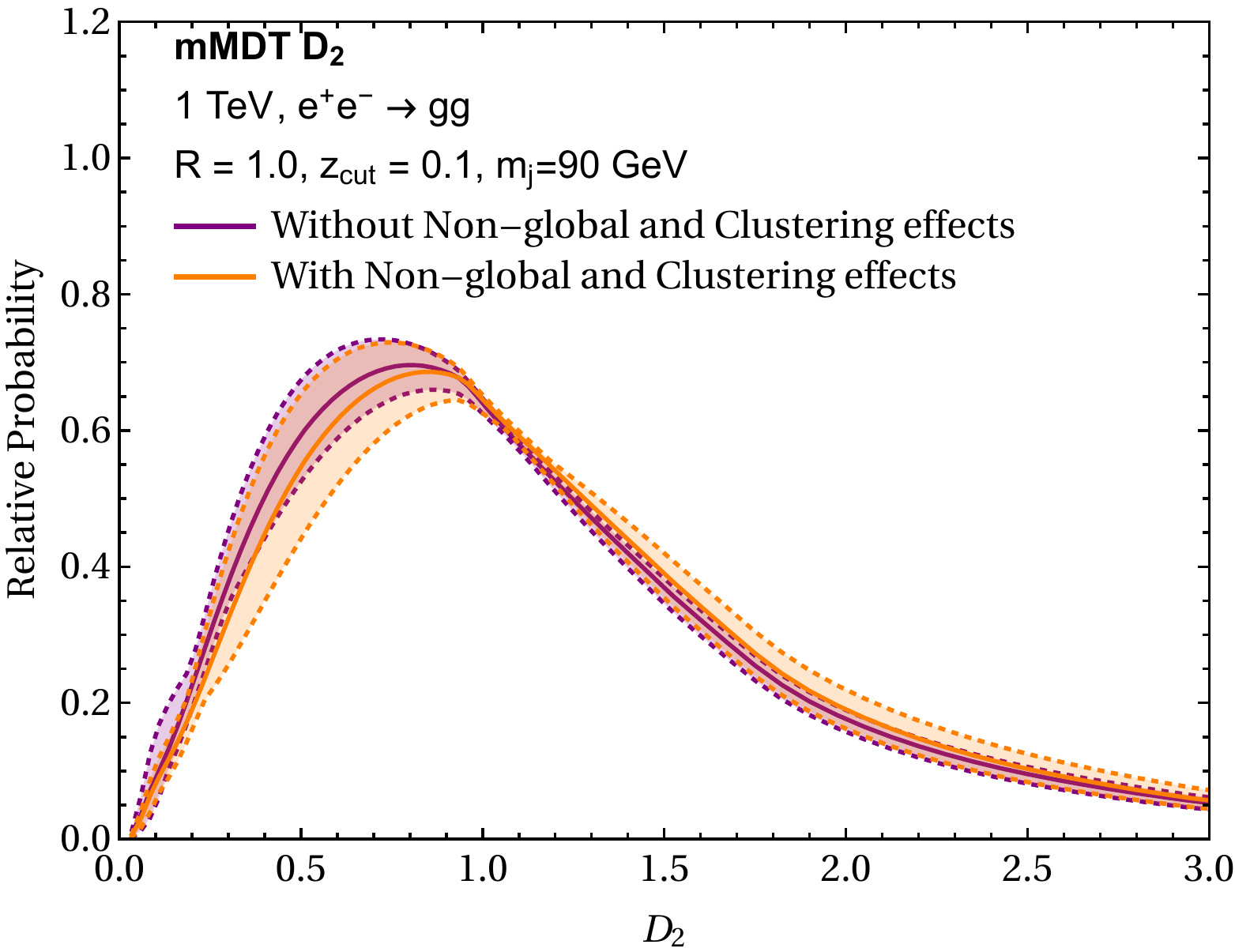}\\
\includegraphics[scale=0.45]{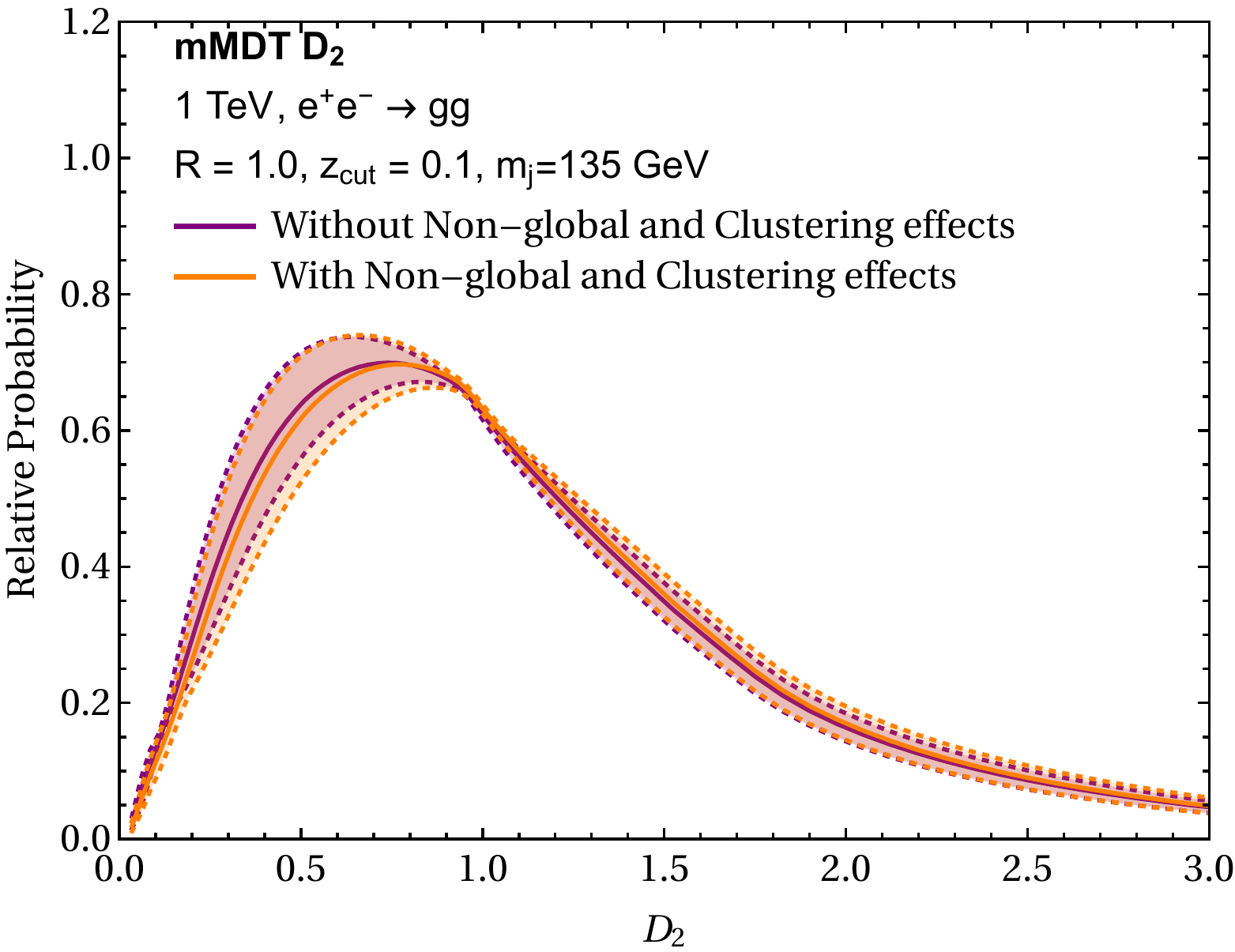}
\caption{\label{fig:spectra_with_NGLs} The $D_2$ relative probability spectra, for a variety of groomed jet masses for gluon initiated jets, and comparing the change in the spectrum with the inclusion of non-global and clustering effects. }
\end{figure}

First we present the NGL/clustering contribution to the collinear-soft function, having factored out the global evolution (this is the direct output of the MC algorithm above):\footnote{The factoring of the non-global contributions from the global is strictly true at leading log accuracy in the non-global logarithms, or equivalently at NLL for the  global logarithms. At higher orders, one would need to perform a convolution between the non-global resummation factor and the solutions to the global RG equations.}
\begin{align}\label{eq:collinear_soft_function}
C_s\Big(e_{3},z_{\text{cut}},\mu\Big)&= U_{\theta<\theta_{ab}}(\mu_{e_3},\mu)U_{\theta>\theta_{ab}}(\mu_{z_{\text{cut}}},\mu) g^{NGL}_{SD}(t,\theta_{ab})\,,\\
  t&=\frac{C_A}{\pi}\int_{\mu_{e_3}}^{\mu_{z_{\text{cut}}}}\frac{d\mu'}{\mu'}\alpha_s(\mu')\,,\\
  \mu_{e_3}&=\frac{QD_2}{2}\sqrt{e_{2}\Big(z(1-z)\Big)^3}\\
  \mu_{\zcut}&=\frac{Q\zcut}{2}\sqrt{\frac{e_{2}}{z(1-z)}}
\end{align}
The scales $\mu_{e_3}$ and $\mu_{z_{\text{cut}}}$ minimize the logarithms given by the calculation of the one-loop anomalous dimensions, as discussed in \Ref{Larkoski:2017cqq}. To calculate the cross-section with non-global and clustering effects, we laplace transform the cross-section of Eq. \eqref{eq:factorized_cross_section}, solve the renormalization group equations in laplace space for each function with generic scales, all evolved to a common scale $\mu$. We then invert the laplace transform analtyically, and take the cumulative distribution up to some maximum $D_2$. We then fix the scales in the cumulative resummed distribution. This procedure resums all global logarithms to NLL accuracy, and to add the clustering and the non-global effects, we multiply this cumulative resummed distribution by $g^{NGL}_{SD}$. Taking the derivative will give the differential resummed cross-section, as plotted in Figs. \ref{fig:spectra_with_NGLs}. We use an angular cutoff of $\delta=0.002$ for the shower in what follows.

\begin{figure}\centering
\includegraphics[scale=0.45]{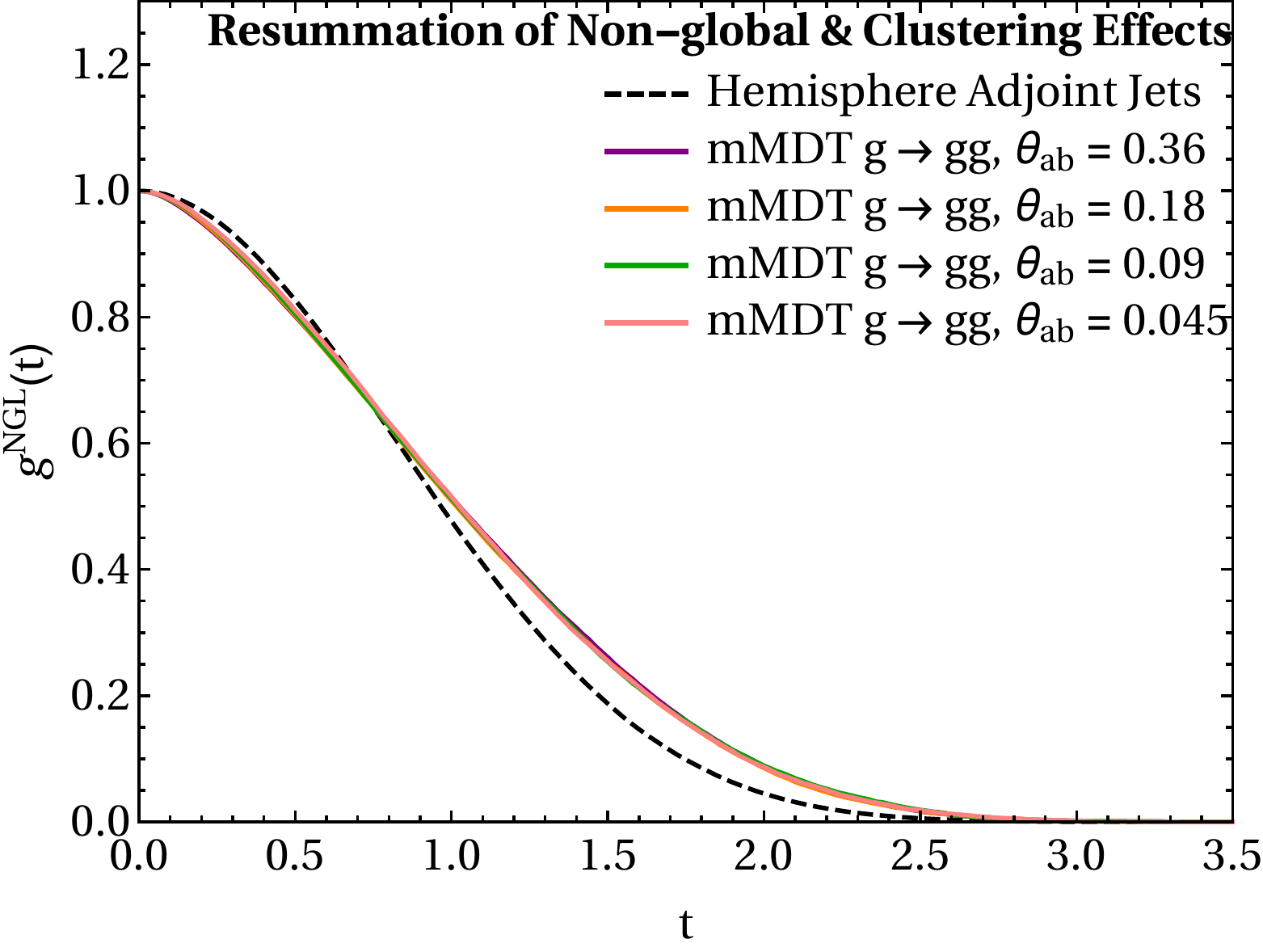} 
\includegraphics[scale=0.45]{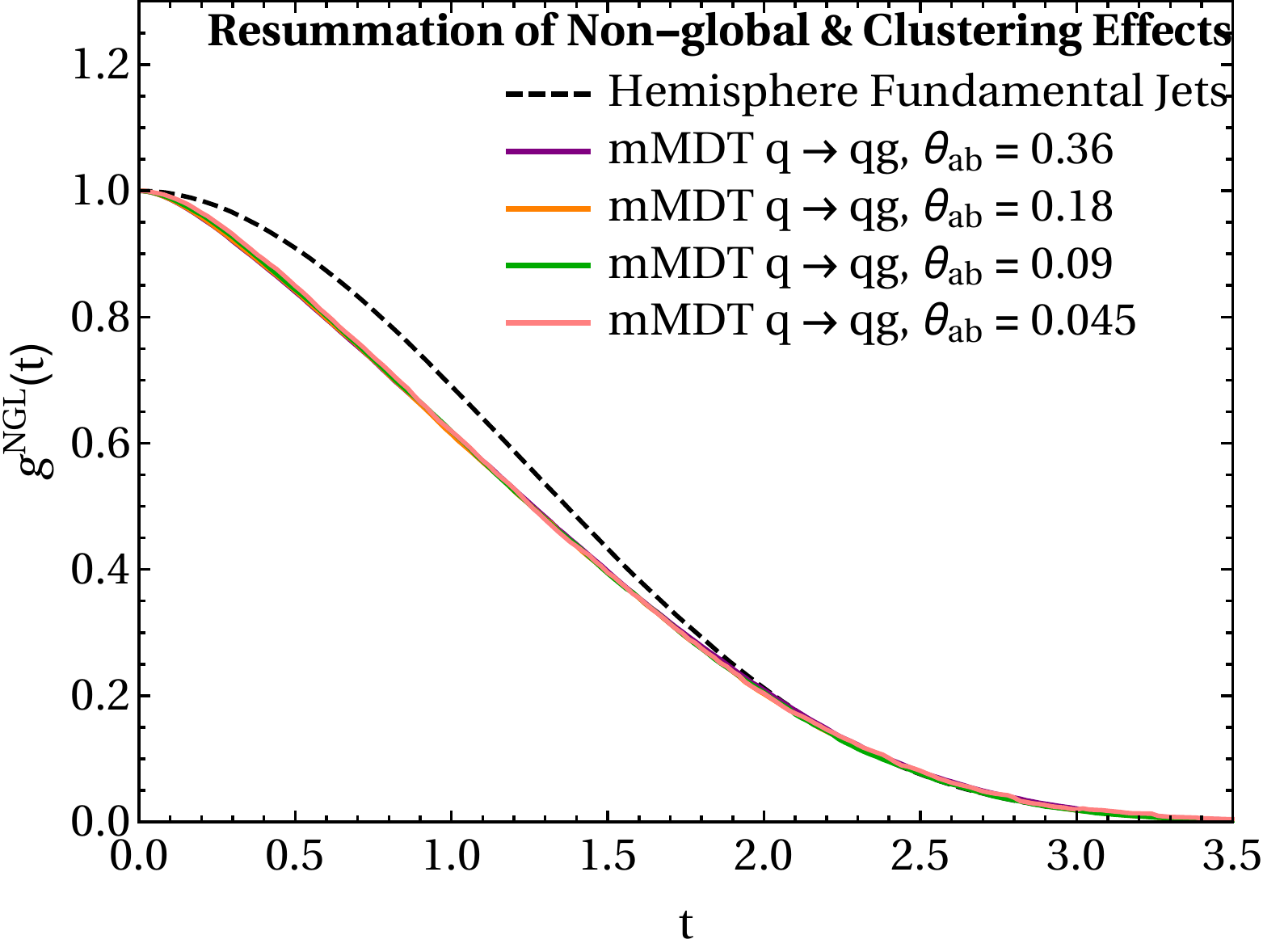}\\
\includegraphics[scale=0.45]{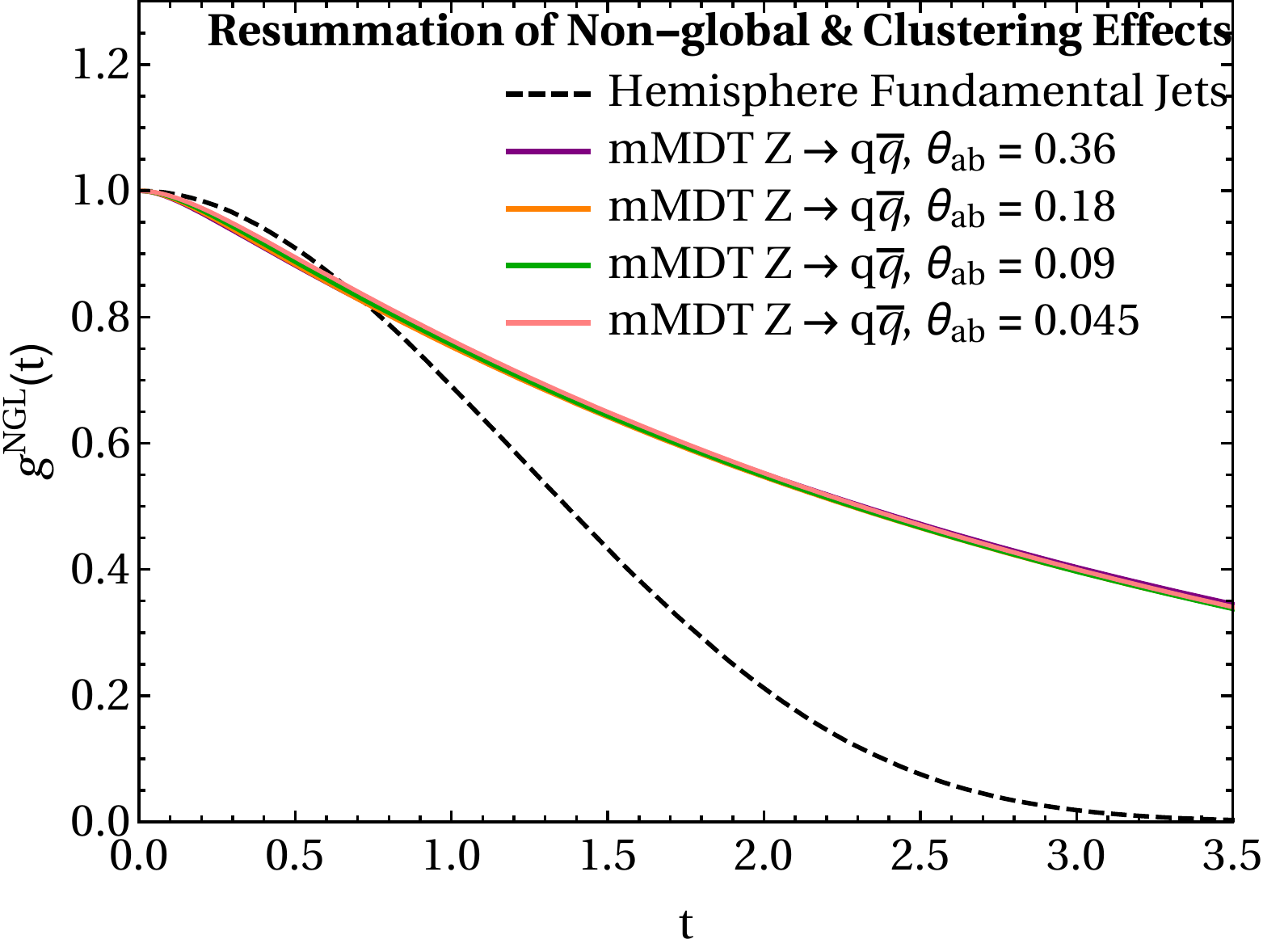}
\caption{\label{fig:NGL_distribution} The resummation of non-global and clustering effects for modified mass drop, with a variety of opening angles and flavor structures, as compared to the classical dijet (either $e^+e^-\rightarrow gg$ or $e^+e^-\rightarrow q\bar{q}$) hemisphere invariant mass NGL distribution. }
\end{figure}
\begin{figure}\centering
\includegraphics[scale=0.45]{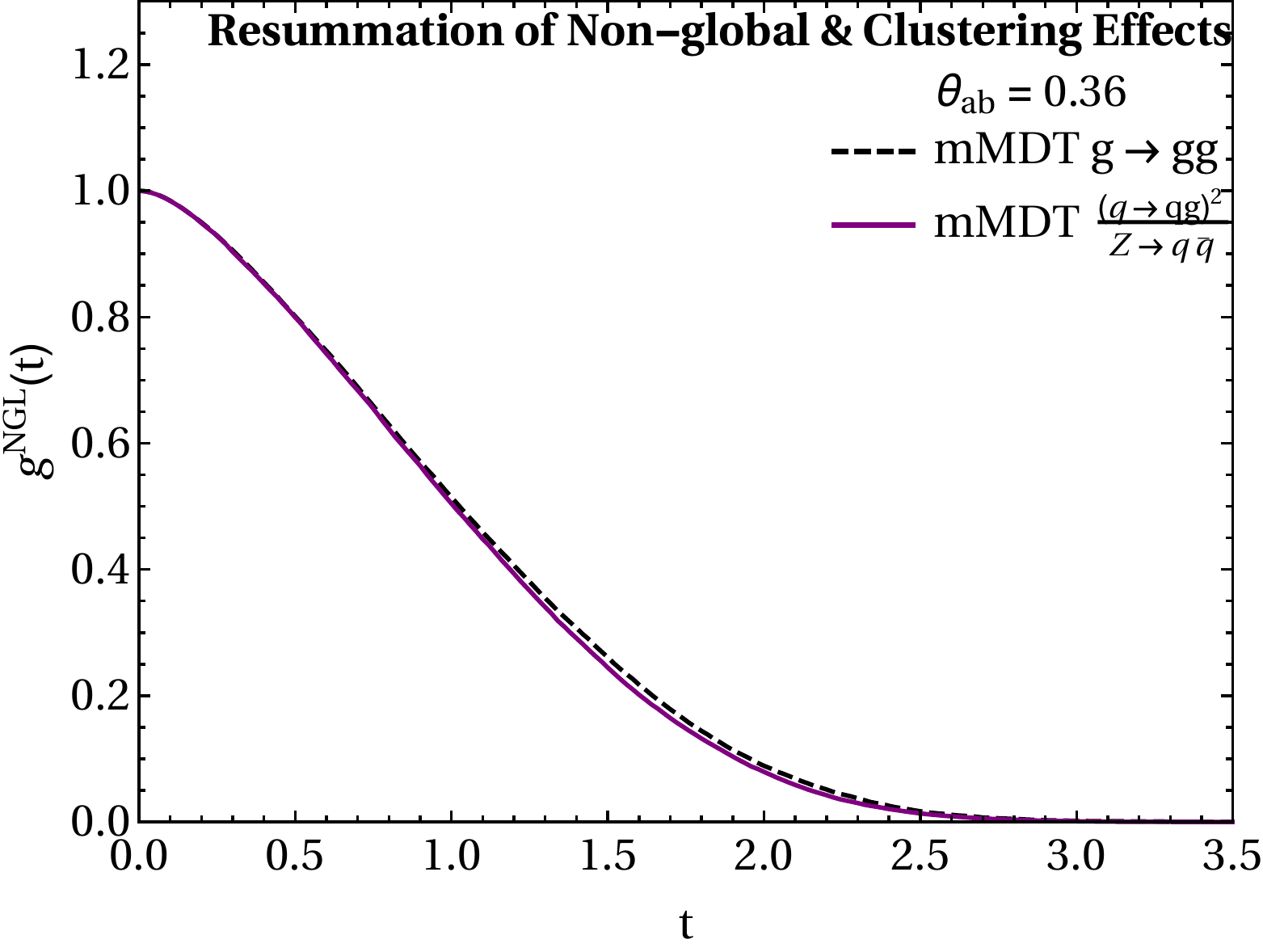}
\caption{\label{fig:dipoles_compare} Testing the large $N_c$ flavor relations, naively violated due to the C/A clustering history phase-space constraints. }
\end{figure}

In Figure \ref{fig:NGL_distribution}, we plot the non-global and clustering modification factor $ g^{NGL}_{SD}(t,\theta_{ab})$, for the splittings $g\rightarrow gg, q\rightarrow qg,\text{ and } Z\rightarrow q\bar{q}$. We give the distribution for a variety of opening angles $\theta_{ab}$ for the collinear splitting, and find that as $\theta_{ab}\rightarrow 0$, the distribution tends to a universal value, very weakly dependent upon the exact value of $\theta_{ab}$. The fact that the quark initiated splittings tend to the same asymptotic value is expected given the arguments of \Ref{Neill:2016stq}. The asymptotics is determined by the number of legs in the active jet region, which in this case is the one recoil direction. Moreover, we find in Figure \ref{fig:dipoles_compare} that we have to a good approximation the different flavor splittings satisfy:
\begin{align}
 g^{NGL}_{SD}(t,\theta_{ab};g\rightarrow gg)\approx\frac{ \Big(g^{NGL}_{SD}(t,\theta_{ab};q\rightarrow qg)\Big)^2}{g^{NGL}_{SD}(t,\theta_{ab};Z\rightarrow q\bar{q})}\,.
\end{align}
This is a very unexpected result, since the different initial dipole configurations ought to lead to very different branching histories, which the C/A clustering is sensitive to. If the real emission phase space constraint did not depend upon the emission history off of all dipoles, like in the hemisphere case (where the geometrical constraint for real emissions is the same for all soft emissions to all orders), such a result would have been expected, based on the large $N_c$ factorization of color-disconnected dipoles.

\begin{figure}\centering
\includegraphics[scale=0.45]{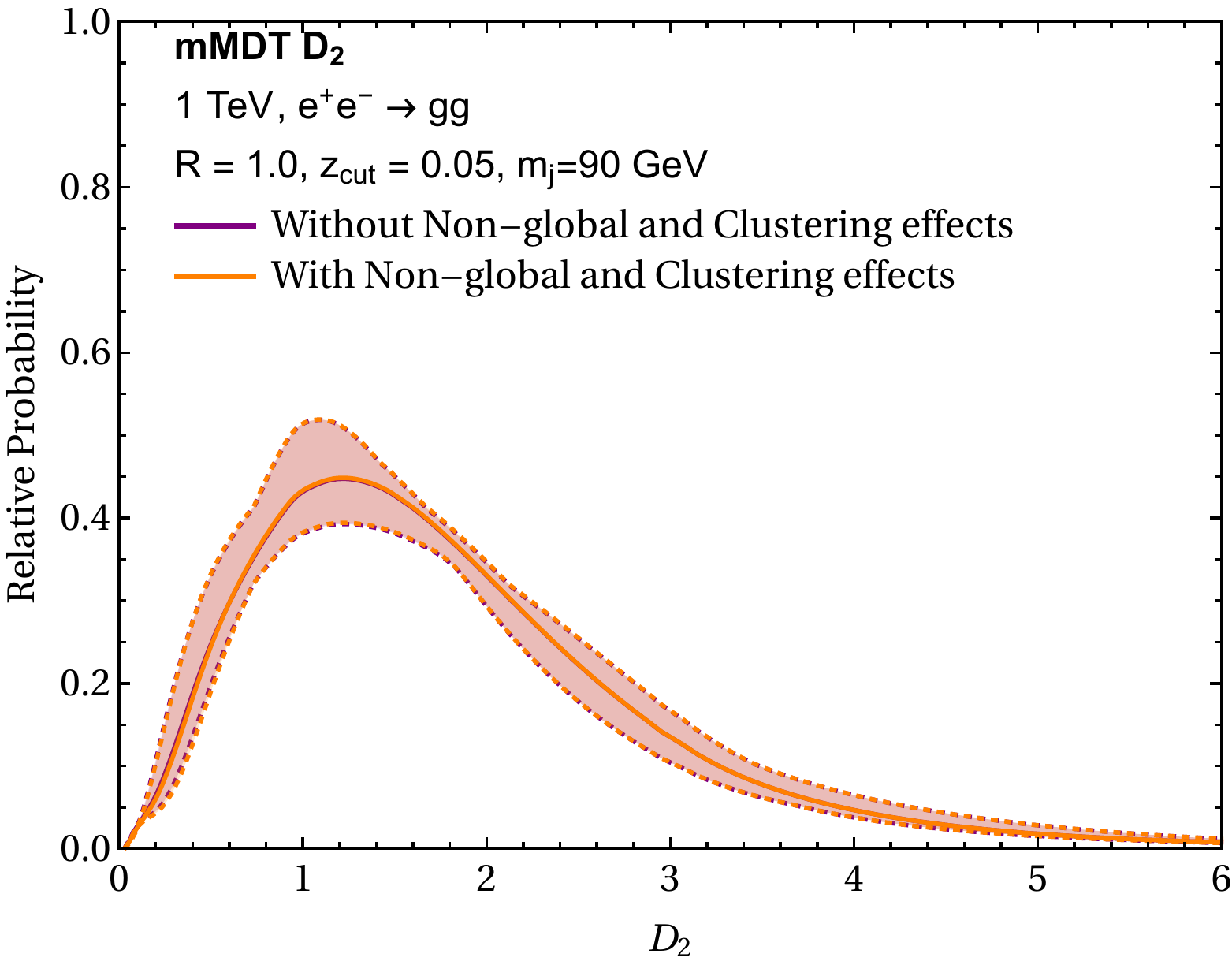} 
\includegraphics[scale=0.45]{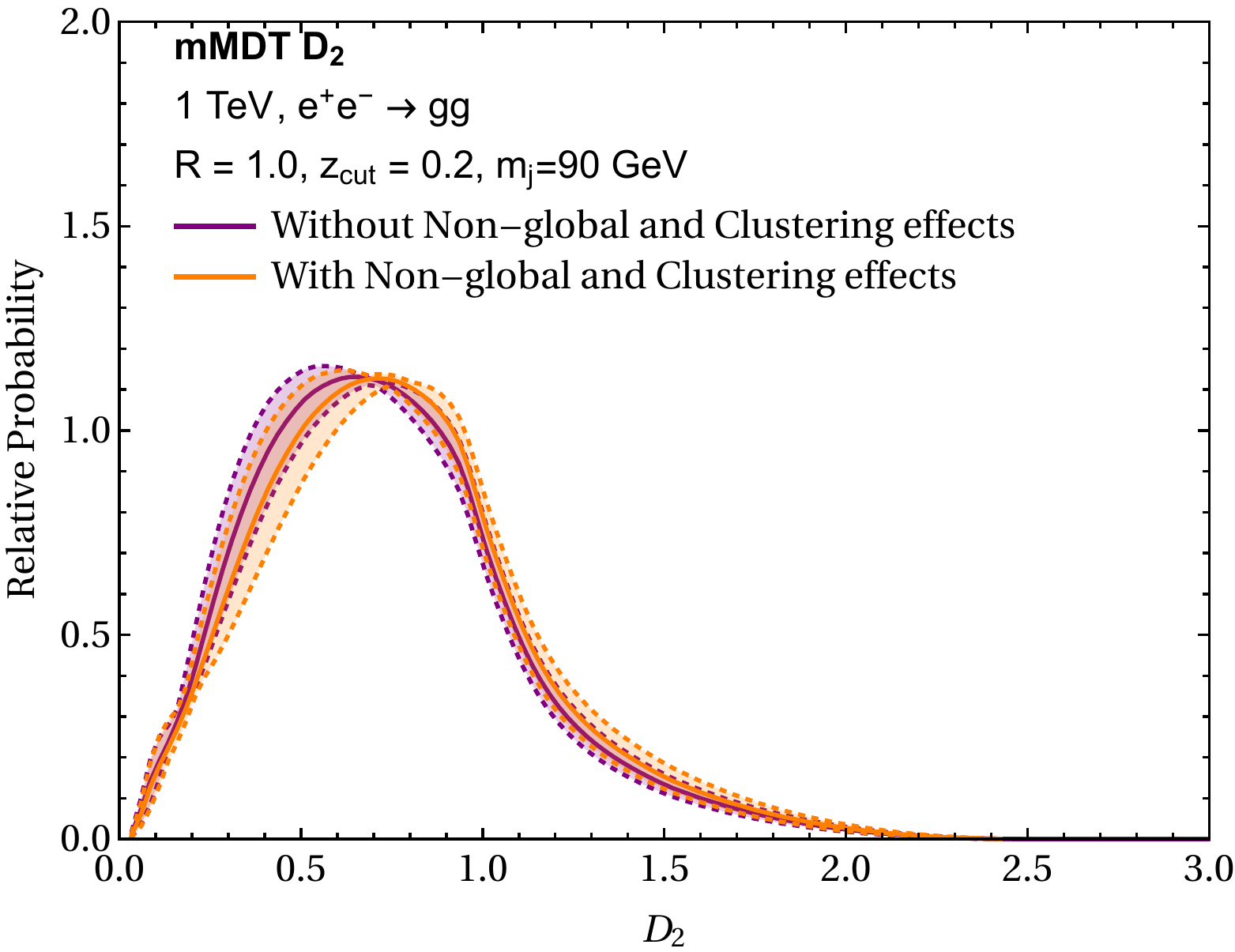}
\caption{\label{fig:zcut_vary} The $D_2$ relative probability spectra, for a variety of groomed jet masses for gluon initiated jets, and comparing the change in the spectrum with the inclusion of non-global and clustering effects. We give the effect on the distribution for gluon jets for $z_{\text{cut}}=0.05$ and $0.2$ for $m_J=90 GeV$. }
\end{figure}

In Figure \ref{fig:spectra_with_NGLs}, we plot the difference between the $D_2$ spectra with non-global and clustering effects, and the spectra with simple exponentiation of the global anomalous dimensions. We give the results for gluon initiated jets, with three different mass cuts $m_j=45, 90$, and $135$  GeV, with a jet energy of $500$ GeV, in order to probe different opening angles. Since the non-global and clustering distribution is strongest for the gluon, we do not include distributions for the quark or $Z$ initiated jets. We can clearly see that the non-global and clustering effects are well within the uncertainty estimates due to scale variation of the starting and ending scales of resummation (this is including variations in where to start the non-global resummation), and that for the most part, simply exponentiating the global anomalous dimensions gives an accurate description of the NLL spectrum. Ultimately, this small effect of the non-global and clustering logarithms is due to the ratio of scales of the distinct soft regions in the function $C_s$ (see Appendix A of \Ref{Larkoski:2017cqq}):
\begin{align}
\frac{\mu_{z_\text{cut}}}{\mu_{e_3}}=\frac{z_{\text{cut}}}{D_2 z(1-z)}\,.
\end{align}
$z$ is the energy fraction of one of the subjets of the splitting. We note that this is a pessimistic estimate for the ratio of scales the non-global resummation is sensitive to, since:
\begin{align}
\frac{z_{\text{cut}}}{D_2 z(1-z)}>\frac{z_{\text{cut}}}{D_2}\,,\forall z\in [z_{\text{cut}},1-z_{\text{cut}}]\,.
\end{align}
For $z_\text{cut}= 0.1$ or $0.05$, this is never a very large ratio of scales (that is, much much greater than 1) until well after the sudakov suppression of the cross-section sets in. We illustrate the effect of changing $z_{\text{cut}}$ in Fig. \ref{fig:zcut_vary}. However, we caution that as $\zcut\rightarrow 0$, non-global effects associated with power corrections due to the expansion $\frac{m_J^2}{E_J^2}\ll \zcut $ can become important, and which are not considered in this study. Note that these non-global effects would also effect the soft-drop/mMDT groomed jet mass distributions of Refs. \cite{Dasgupta:2013ihk,Dasgupta:2013via,Frye:2016aiz,Marzani:2017mva}.

\begin{figure}\centering
\includegraphics[scale=0.35]{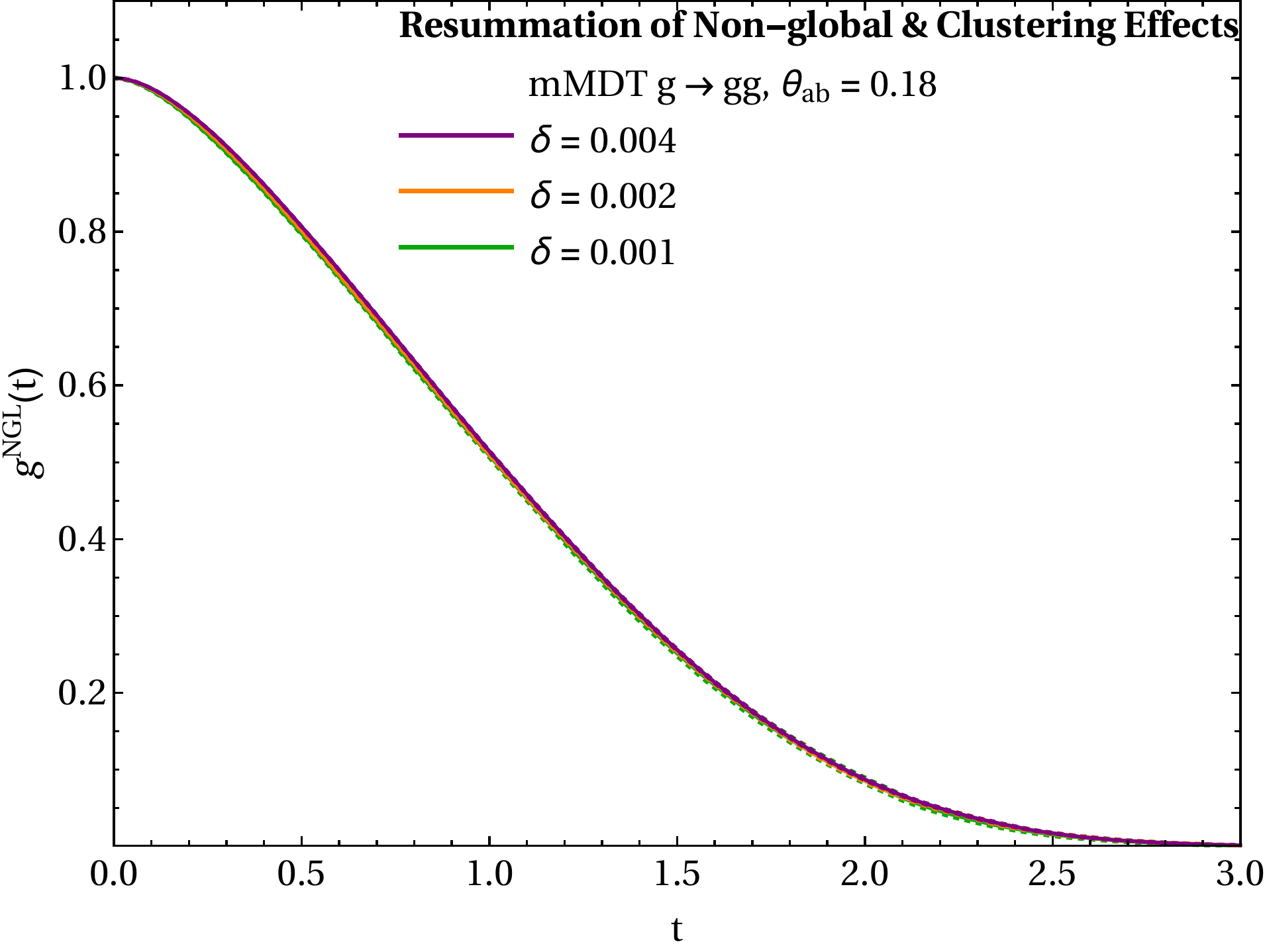} 
\includegraphics[scale=0.35]{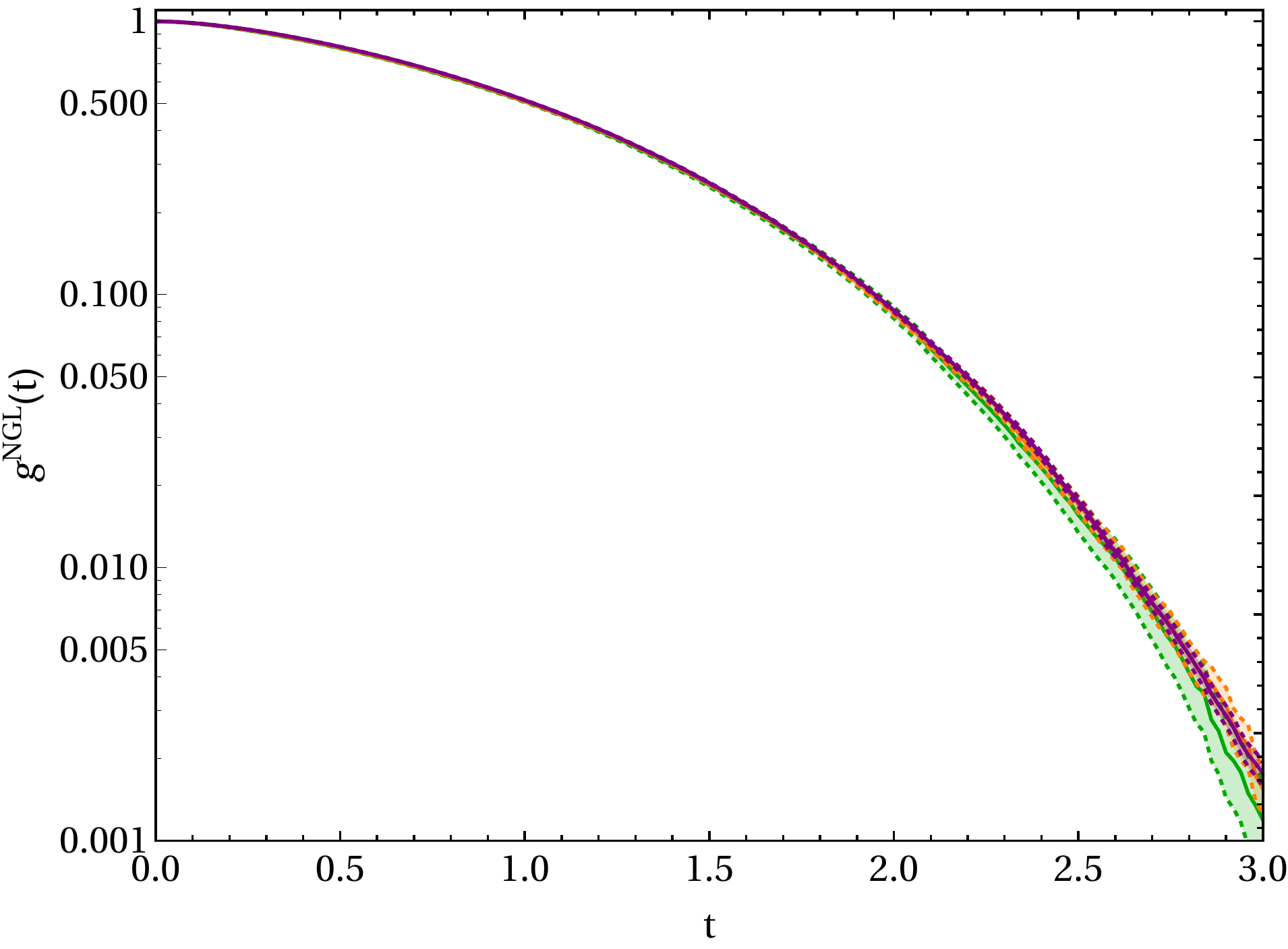}
\caption{\label{fig:converge} The cutoff dependence and statistical uncertainties for the MC determination of the NGL spectrum. Shown are gluon initiated jets with angular cutoffs $\delta=0.001, 0.002$ and $0.004$. The bands (present in both plots) represent the statistical uncertainty, and the spectra are shown for both linear and logarithmic scales.}
\end{figure}
Finally, in Fig. \ref{fig:converge}, we show the dependence on the angular cutoff of the NGL MC for gluon initiated jets, as well as the statistical uncertainties. The bands represent the root mean squared (RMS) fluctuations for a MC run containing approximately 50,000 events for $\delta=0.001$, 250,000 events for  $\delta=0.002$, and  5000,000 events for $\delta=0.004$. The RMS spread was estimated by running the MC for 10 statistically independent runs, each containing the same number of events. We then calculate the RMS over the 10 runs. Finally we smooth the error by fitting an exponential function $a\times e^{b t}$ to the ratio of the RMS to the mean as a function of $t$. We then take the upper and lower bounds as given by $g^{NGL}_{\pm}(t)=g^{NGL}_{ave.}(t)(1\pm a\times e^{b t})$ as the estimation of the statistical uncertainity. We find negligible statistical uncertainty and cutoff dependence for NGL values out to $t\sim 2.5$.

\section{Conclusions}
We have investigated the contribution of non-global and clustering effects that directly change the shape of the groomed shape variable $D_2$ for multi-pronged jets using the mMDT grooming procedure. In all, there are three sources of non-global contributions to the groomed $D_2$ observable. Two of them are directly shared with the groomed mass distribution of Refs. \cite{Dasgupta:2013ihk,Dasgupta:2013via,Frye:2016aiz,Marzani:2017mva}. First, there are the non-global contributions directly contributing to the groomed constituents of the jet, which for an ungroomed jet would modify its mass spectrum. At leading power, these contributions are removed, a major feature of the mMDT algorithm, and effectively erasing any hard geometric boundary of the jet with respect to the rest of the event, so that the jet appears to have zero active area. Secondly, there are non-global contributions which can change the relative quark and gluon jet fractions. These contributions correspond to so-called ``global'' soft modes which are not associated to any particular jet, and are therefore sensitive to the precise jet boundaries drawn over the event, as well as the additional cuts one places on the event. These contributions cannot directly affect the shape of a quark jet or a gluon jet mass or $D_2$ spectrum (the global soft functions being identical in the two cases). Thus for $e^+e^-$ collisions, such contributions can be normalized away, since jets are dominated by the quark initiated process, but may play a role in hadron-hadron jet spectra. In the hadron-hadron collision case, they can also be resummed in each quark and gluon fraction for fat jets using the techniques of \Ref{Dasgupta:2012hg}. Finally, there are the contributions to the $D_2$ spectrum which do not change the mass spectrum of the groomed jet. These contributions are what were considered and resummed in this paper. Again, these contributions are not sensitive to the precise jet boundary for large $R$ jets, and are determined by specifying the flavor structure of the boosted jet. The secondary branching takes place outside of the region defined by the opening angle of the dipole required to exist in the $D_2\rightarrow 0$ limit, and whether the branching is inside or outside the groomed jet is irrelevant as all such radiation fails the grooming requirement.

\begin{acknowledgments}
 D.N. acknowledges helpful discussions with Ian Moult and Andrew Larkoski, and thanks both the Berkeley Center for Theoretical Physics and LBNL for hospitality while portions of this work were completed, as well as support from the Munich Institute for Astro- and Particle Physics (MIAPP) of the DFG cluster of excellence ``Origin and Structure of the Universe'', and support from DOE contract DE-AC52-06NA25396 at LANL and through the LANL/LDRD Program.
\end{acknowledgments}

\bibliography{reply_bib.bib}
\end{document}